\documentclass[11pt]{pazh}

\usepackage{psfig}

\usepackage[T2A]{fontenc}
\usepackage[utf8]{inputenc}

\usepackage{latexsym}
\usepackage{amssymb}

\begin{document}

\setcounter{page}{1}

\sloppypar

\title{\bf Statistics of interacting galaxies at $z \sim 0.7$}

\author{V.P. Reshetnikov\inst{1,2}, Y.H. Mohamed\inst{1,3}} 

\institute{St.Petersburg State University, Universitetskii pr. 28, Petrodvoretz, 
198504 Russia
\and
Observatoire de Paris, LERMA, CNRS, 61 av. de l'Observatoire, Paris, F-75014, France
\and
Astronomy Department, National Research Insitute of Astronomy and Geophysics,
Cairo 11421, Egypt
}

\titlerunning{Interacting galaxies}

\abstract{
We present the results of our analysis of the frequencies of galaxies 
with tidal tails and M51-type galaxies in several deep fields of the Hubble 
Space Telescope (HDF-N, HDF-S, HUDF, GOODS, GEMS). In total, we have found about 
seven hundred interacting galaxies at redshifts $z \leq 1.5$ in these
fields. At $z \leq 0.7$, the observed space densities of galaxies with tidal structures and M51-type galaxies
have been found to increase as  $\propto (1+z)^m$, where $m \approx 2.6$.
According to our estimates, over the last 6-7 Gyr, i.e., at $z \leq 0.7$,
about a third of the galaxies with $M(B) \leq -18^m$ must have undergone strong
gravitational perturbations and mergers and $\sim$1/10--1/5 
of the galaxies have accreted relatively low-mass
nearby satellites typical of M51-type galaxies. The possible decrease in 
the timescale on which a distant galaxy appears peculiar with growing 
$z$ can increase considerably the estimated rate of mergers.
}
\titlerunning{Interacting galaxies}
\maketitle

\section{Introduction}

The idea about the growth of stellar systems inside
hierarchically merging dark haloes provides a basis
for the present-day models of galaxy formation and
evolution (see, e.g., White and Rees 1978; for a review,
see the book by Mo et al. 2010). The process of
halo mergers can be well described in terms of model
calculations (Fakhouri and Ma 2008 and references
therein); however, the connection of this process with
the mergers of actually observed galaxies is not yet
understood well enough (see, e.g., Kitzbichler and
White 2008).

As a rule, two approaches are used when the
galaxy merger rate is studied observationally. The first
is based on analysis of the frequency of close galaxy
pairs at various $z$ (Zepf and Koo 1989; Kartaltepe
et al. 2007; de Ravel et al. 2011; and references
therein). The weak points of this method are the small
number of spectroscopically confirmed pairs at high $z$
and the assumption that all these pairs merge on a
certain timescale (Kitzbichler and White 2008).

The second (morphological) approach is based on
the frequency statistics of signatures of recent interactions
and mergers in galaxies: structure distortions
(Conselice et al. 2003 and references therein),
the presence of tidal tails and bridges (Reshetnikov
2000a, 2000b; Bridge et al. 2010), polar rings
(Reshetnikov 1997; Reshetnikov and Dettmar 2007),
collisional rings (Lavery et al. 2004), etc. The main
problem of this method is that the basic tracers of
dynamical perturbations of galaxies, as a rule, have
a low surface brightness and they are difficult to observe
at high $z$. However, as was shown by Hibbard
and Vacca (1997), such structures must be visible
at least up to $z \sim 1$ on deep images of the Hubble
Space Telescope (HST). The search for and statistics
of tidal structures in galaxies in the HDF-N
and HDF-S deep fields confirmed this conclusion
(Reshetnikov 2000a, 2000b). Furthermore, as in
the case of studying merging pairs, the connection
of the dark halo merger rate derived from models
with the complex processes of galaxy interactions and
mergers acting on different mass and timescales is
ambiguous. On the other hand, the advantage of the
morphological approach is that very large samples of
galaxies can be produced and investigated using it.

The results of the two approaches do not yet agree
quite well quantitatively; however, the general conclusion
is beyond question: the fraction of interacting
and merging galaxies actually grows with $z$, as follows
from theoretical expectations. At $z \leq 1.5$,
this growth is commonly described by a power law,
$(1 + z)^m$, where the value of $m$ from the data of different
authors is, in most cases, within the range from 2 to 4
(see, e.g., Table~2 in Kartaltepe et al. 2007). 

In this paper, we analyze the frequency statistics of
galaxies with tidal structures and M51-type galaxies in
several Hubble deep fields. All numerical values in
the paper are given for the cosmological model with a
Hubble constant of 70 km s$^{-1}$ Mpc$^{-1}$ and 
$\Omega_m=0.3$, $\Omega_{\Lambda}=0.7$.

\section{Galaxies with tidal structures at $z \sim 0.7$}

\subsection{The sample of galaxies}

The sample is based on the catalog of interacting
galaxies in several Hubble deep fields (Mohamed
and Reshetnikov 2011). This catalog was compiled
on the basis of a visual classification of galaxies in
the F814W (HDF-N and HDF-S fields), F775W (HUDF), and 
F850LP (GOODS and GEMS) filters. At $z \sim 1$,
these filters roughly correspond to
the $B$ band in the reference frame associated with
the galaxies themselves. Our sample includes a total
of 689 galaxies with tidal tails (the bridges have, on
average, a lower surface brightness and they were
not considered in our work) at redshifts $z \leq 1.5$.

\begin{figure}
\centerline{\psfig{file=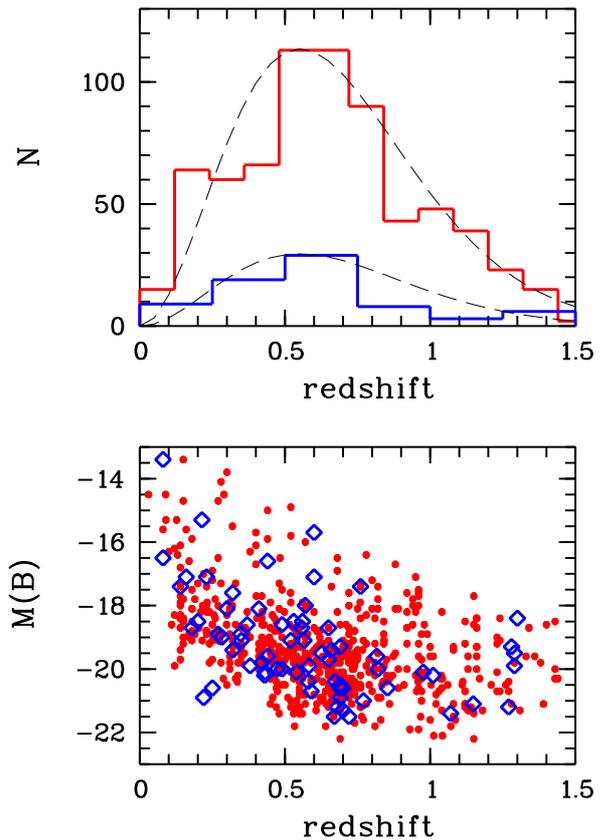,width=8cm,angle=-90,clip=}}
\caption{Top: Redshift distributions of the samples of galaxies with 
tidal tails (red histogram) and M51-type objects (blue
histogram). The dashed lines indicate the expected distributions 
for $z_m = 0.64$. 
Bottom: The absolute magnitude -- redshift relation for galaxies with 
tidal tails (red circles) and M51-type galaxies (blue diamonds).}
\end{figure} 

Figure~1 shows the $z$ distribution of sample galaxies
and their positions on the ``rest-frame absolute magnitude --
redshift'' diagram. The dashed lines in the upper panel show
the fits of the observed distributions by the empirical formula
$dN/dz \propto z^2\,{\rm exp}(-(z/z_c)^{1.5}$
proposed for describing the observational data for magnitude-limited 
samples with median redshift $z_m = 1.412 z_c$
(Baugh and Efstathiou 1993). As we see from the figure, the
actual distributions agrees satisfactorily with the expected
ones.

Figure~1 (bottom part) illustrates the obvious observational selection
effect: we predominantly choose the brightest
galaxies among more distant objects. The standard
way of avoiding this selection is to consider only
bright galaxies at various $z$. Below, we will study the
statistics of interacting galaxies for two subsamples:
(1) for objects with absolute $B$ magnitude from
--18$^m$ to --20$^m$ and 
(2) for all galaxies with
$M(B) \leq$--18$^m$ (the stellar mass limit
M$\geq 4 \cdot 10^{9}$\,M$_{\odot}$ roughly
corresponds to this luminosity restriction).

To estimate the completeness of our sample,
we considered the differential counts of galaxies in
comparison with those in the VIRMOS VLT F02
deep field (McCracken et al. 2003). According to
McCracken et al. (2003), the differential counts for
galaxies brighter than 24$^m$ in the $I$ band have a
slope of 0.34$\pm$0.02. (This result agrees well with
the data on other deep fields as well; e.g., Metcalfe
et al. 2001). The objects in our sample of interacting
galaxies follow this slope up to $I \approx 21.5$ and then 
begin to deviate greatly from it. The median redshift
for galaxies with $I$ = 21.5 is $z$ = 0.67 (this value is
also close to that found from the analytical fit shown
in Fig.\,1) and below we take precisely this redshift
as a completeness limit for the sample of interacting
galaxies (both for galaxies with tidal tails and M51-type
objects).

\subsection{Estimating the evolution rate}

To study the evolution of the frequency of galaxies
with tidal structures, we used the same approach as
that in Reshetnikov (2000a, 2000b). The meaning of
this method is that, having fixed the space density of
objects of some type at $z = 0$, the expected number
of such galaxies is estimated within the selected field
in a given $z$ range for different space density evolution
laws. By comparing the actual and expected numbers
of objects, we can obtain a constraint on the density
evolution law. Below, we assume that the space density of
galaxies varies with $z$ as $n(z) = n_0 \times (1 + z)^m$
and estimate the exponent $m$.

An important stage in estimating the evolution
rate is to determine the local density of galaxies
$n_0$. Unfortunately, for galaxies with tidal structures, this
quantity is known relatively poorly. In addition, it
depends on the brightness level at which the structure
is distinguished: for example, at a brightness level of
$\sim28^m/\Box''$, more than 10\% of spiral galaxies
exhibit various kinds of external structures (Miskolczi
et al. 2011). In our paper, we used the results of Nair
and Abraham (2010), who performed a detailed visual
classification of approximately 14\,000 galaxies from
the Sloan Digital Sky Survey; they also provided the
frequency statistics of galaxies with tidal tails (see Table
4 in their paper). According to Nair and Abraham,
the visually distinguishable tidal tails are seen in 301
of 14\,034 galaxies, i.e., in about 2\% of the galaxies.
As the luminosity function of nearby galaxies, we take
the results of the 2dF survey (Norberg et al. 2002) and
assume that the fraction of galaxies with tidal tails in
any luminosity range is 0.02.

When nearby and distant galaxies are compared,
the possible evolution of their luminosity with $z$
should be taken into account. Studies of the Tully--Fisher
relation and the luminosity function of distant
galaxies show that the galaxies at $z \sim 1$
were approximately 1$^m$ brighter (see, e.g., Gabasch et al. 2004;
Miller et al. 2011). Therefore, to properly compare the
space densities of galaxies at different $z$, this effect
should be taken into account. As a model for the
luminosity evolution of spiral galaxies, we used the
results of Bicker et al. (2004). In the redshift range
of interest ($z < 0.7$), the predictions of the model by
Bicker et al. for the luminosity evolution of
Sb--Sc spirals ($\Delta M(B) \propto (1^m-1.^m2) \times z$) are close
to the actually observed changes.

Table\,1 presents the final results of determining
the exponent $m$ from the statistics of galaxies with
tidal tails within the deep fields we considered (their
total angular area is $\Omega = 6.83 \times 10^{-5}$ sr)
at $z \leq 0.67$. As the errors in $m$, the table provides the range of
variation in the exponent in the case of a Poissonian
error in the number of objects in the field ($\pm\sqrt{N}$);
the total number of galaxies in a given luminosity
range is given in parentheses ($N$); the first and second
rows give the value of $m$ in the absence of luminosity
evolution (only the $k$-correction was applied to the
absolute magnitudes) and with its allowance. As
we see from the table, allowance for the luminosity
evolution changes the $m$ estimates, but not too much,
approximately by 10\%.

\begin{table}
\caption{Results of the determination of the parameter $m$}
\begin{center}
\begin{tabular}{|c|c|c|}
\hline
Sample & $-18^m \geq M_B \geq -20^m$ & $M_B \leq -18^m $  
\\
\hline               
               &                &              \\
Galaxies    & 2.56$^{+0.17}_{-0.18}$~~(198) & 2.93$^{+0.14}_{-0.15}$~~(278)    \\
  with tidal tails     &  2.71$^{+0.16}_{-0.17}$~~(211)  & 2.60$^{+0.15}_{-0.16}$~~(243)  \\
               & &      \\
  M51-type                &   2.97$^{+0.43}_{-0.52}$~~~(27) & 3.13$^{+0.37}_{-0.46}$~~~(34)  \\
  galaxies          &   2.69$^{+0.45}_{-0.56}$~~~(24)   & 2.48$^{+0.43}_{-0.54}$~~~(26)  \\
        & & \\
\hline
\end{tabular}
\end{center}
\end{table}

\section{The frequency of M51-type galaxies}

M51-type galaxies are binary systems that consist,
as a rule, of a bright spiral galaxy with a relatively
low-mass satellite located near the end of one of its
spiral arms (Klimanov and Reshetnikov 2001). In
the nearby Universe, such systems are relatively rare
(only about 0.3\% of all galaxies can be attributed to
this type; see Klimanov 2003). However, they are
convenient objects for studying a number of questions
in the physics of galaxies (e.g., the spiral arms formed
by tidal perturbations, the effects of satellites on the
structure and star formation in galaxies).

\begin{figure*}
\centerline{\psfig{file=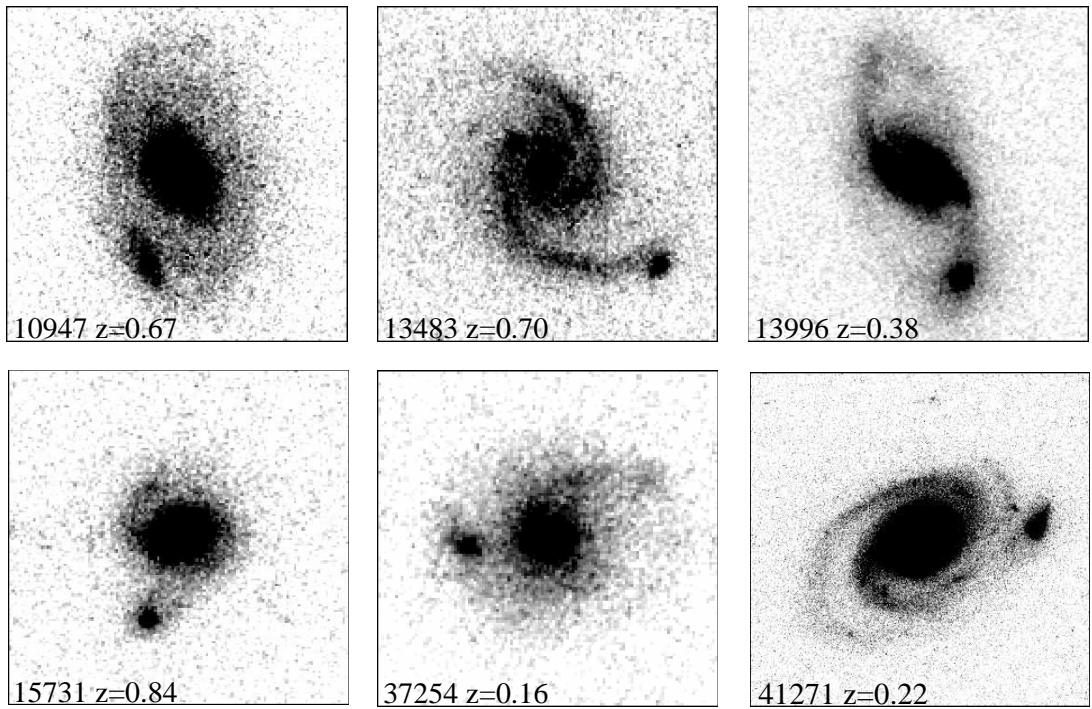,width=15cm,clip=}}
\caption{Examples of M51-type galaxies in the GEMS field. The images were 
extracted from the frames in the F850LP filter. The image sizes 
for the first four galaxies are $5.''4 \times 5.''4$, $3.''6 \times 3.''6$
for 37254, $18'' \times 18''$ for 41271. The galaxy number
and redshift according to the catalog of Wolf et al. (2004) are 
indicated in the lower left corner of each figure.}
\end{figure*} 

As in the case of galaxies with tidal tails, the
sample of M51-type galaxies is based on the catalog
by Mohamed and Reshetnikov (2011); see the
examples in Fig.\,2. The sample includes 74 such
binary systems (the distributions of galaxies in $z$ and
$M(B)$ are shown in Fig.\,1). Figure\,3 compares the
distributions of nearby (32 binary systems; see Klimanov
and Reshetnikov 2001) and distant galaxies
in ratio of the observed luminosity of the satellite to
the luminosity of the main galaxy ($L_s/L_m$) and in
relative distance to the satellite ($R/r$). As we see
from the figure, the distributions are, on the whole,
similar. The corresponding means are
$\langle Ls/Lm \rangle = 0.19 \pm 0.21$, $\langle R/r \rangle = 1.34 \pm 0.48$
(nearby galaxies) and $0.22 \pm 0.18$, $1.26 \pm 0.53$ (distant galaxies).

\begin{figure}
\centerline{\psfig{file=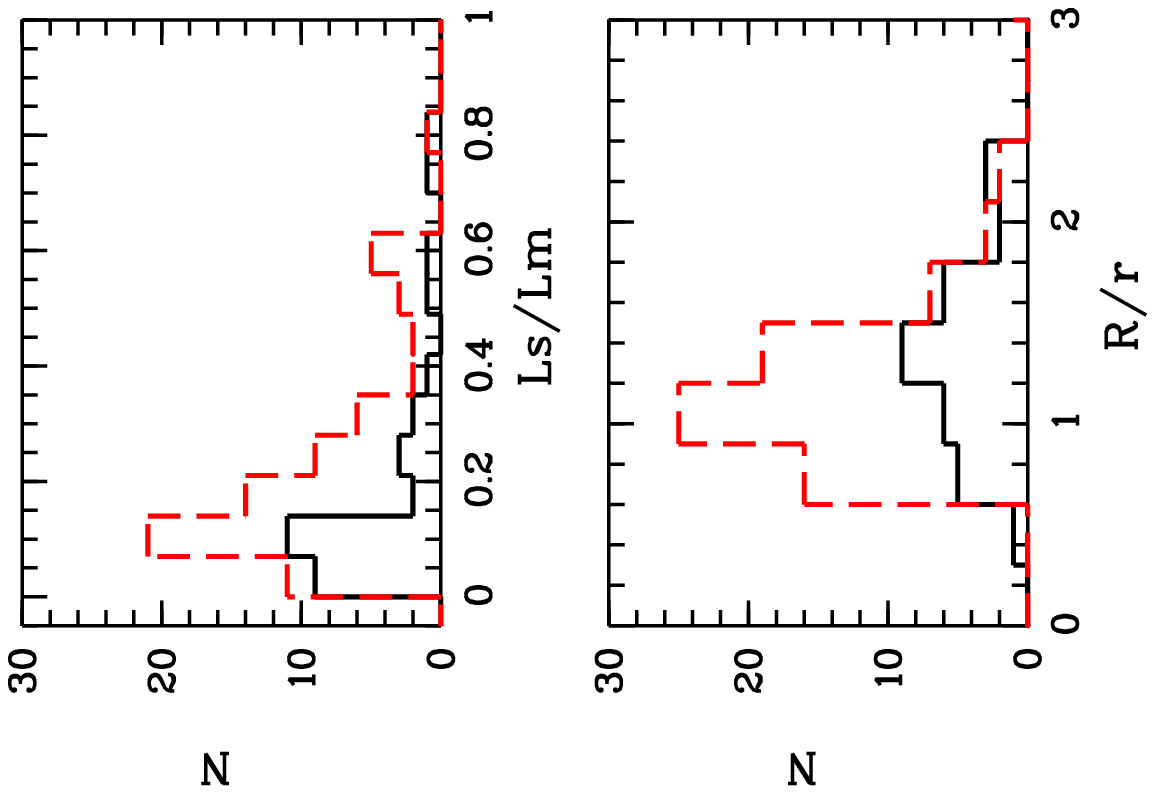,width=8cm,angle=-90,clip=}}
\caption{Top: Ratio of the luminosity of the satellite to the luminosity of 
the main galaxy for the samples of nearby (black solid line) and
distant (red dashed line) M51-type galaxies. Bottom: Ratio of the distance to 
the satellite ($R$) to the radius of the main galaxy ($r$) for
M51-type galaxies (the notation is the same).}
\end{figure} 

The local space density $n_0$ of M51-type galaxies
was found from their luminosity function. According
to Klimanov (2003), the $B$-band luminosity
function can be described by a Schechter function
with the following parameters:
$\phi_*=1.18 \times 10^{-5}$ Mpc$^{-3}$, $\alpha=-1.30$, $M_*(B)=-20.45$.
Then, $n_0[-18^m \geq M(B) \geq -20^m] = 2.51 \times 10^{-5}$ Mpc$^{-3}$
and $n_0[M(B) \leq -18^m] = 2.97 \times 10^{-5}$ Mpc$^{-3}$.
The final estimates of the parameter $m$ for two luminosity
ranges are listed in Table~1. As in the case of galaxies
with tidal tails, the first and second rows give the
estimates without and with allowance for the
luminosity evolution, respectively.

\section{Discussion}

\subsection{Galaxies with tidal tails}

Let us consider how our data agree with the results
of other works. Methodically, the closest work is the
paper by Reshetnikov (2000b), in which it was found
from the statistics of 25 galaxies with tidal tails in
HDF-N and HDF-S that $m=3.6^{+1.2}_{-0.9}$.
To properly compare the results, we should take into account the
fact that Reshetnikov (2000b) assumed the galaxies
with tidal tails to account for 1\% of the field galaxies.
If this estimate is doubled (see Sect. 2.2),
then the $m$ estimate will decrease approximately by a
factor of 1.5 and will be consistent with our results.

The results of several recent works, in which the
morphological approach, i.e., the statistics of signatures
of gravitational perturbations and mergers,
was used to estimate $m$ at
$z \sim 1$ are summarized in Table~2 (see also references
in these papers). The first, second, third, fourth,
fifth, and sixth columns of this table give, respectively,
the references, the type of objects (`m' for
mergers, `rg' for ring galaxies, `ts' for tidal structures, `tt'
for tidal tails), the number of galaxies in the sample
of interacting galaxies (if it is given in the original
paper), the redshift range, the range of $M(B)$ or
another similar characteristic (the infrared luminosity,
the mass of stars), and $m$. Our results on M51-type
galaxies were not included in Table 2, because they
should be compared with the statistics of close pairs.

\begin{table*}
\caption{Values of $m$ from different sources}
\begin{center}
\begin{tabular}{|c|c|c|c|c|c|}
\hline
Reference      & Type of objects & $N$ & $\Delta z$ & $\Delta M(B)$ & $ m$  \\
\hline               
            &              &   &            &               &       \\
Conselice et al. (2003) & m &  & $\leq 1$ & $\leq -18^m$ & 2.5$\pm$0.3  \\
Lavery et al. (2004)& rg & 25& 0.1--1.0&  & 5.2$\pm$0.7 \\
Bridge et al. (2007)& m & 64  & 0.2--1.3 & $L_{IR}>5\cdot10^{10}L_{\odot}$ & 2.12$\pm$0.93 \\
Kampczyk et al. (2007)& m & 68 & $\leq 0.7$ & $\leq -19^m$& 3.8$\pm$1.2  \\
Lotz et al. (2008) & m & 402 & 0.2--1.0& $\leq -18.94-1.3z$ & 0.23$\pm$1.03  \\
Conselice et al. (2009) & m &  & 0.2--1.2 &  $>10^{10}$\,M$_{\odot}$ & 2.3$\pm$0.4 \\
L\'opez-Sanjuan et al. (2009a)& m &25 &0.35--0.85& $\leq -20^m$& 2.9$\pm$0.8 \\
L\'opez-Sanjuan et al. (2009b)& m &61 &0.2--1.1& $\leq -20^m$& 1.8$\pm$0.5 \\
Bridge et al. (2010) & ts & 1075 & 0.25--1.0 & $>10^{9.5}$\,M$_{\odot}$& 2.56$\pm$0.24 \\
This paper&  tt& 243 & $\leq 0.67$ & $\leq -18^m$ & 2.60$^{+0.15}_{-0.16}$\\
       &    tt     & 211 & $\leq 0.67$ & [--18$^m$,--20$^m$] & 2.71$^{0.16}_{-0.17}$\\
       & & & & & \\
                                               \hline
\end{tabular}
\end{center}
\end{table*}
                                               
As we see from Table~2, the works on the statistics
of tidal structures (this paper and the paper by
Bridge et al. 2010) yield very similar results. Bridge
et al. performed a visual classification of about 27\,000
galaxies with $i\leq22.^m2$ in a sky region with an area of
2$\Box^{\rm o}$ (the CFHTLS-Deep project). As a result, they
identified more than a thousand interacting galaxies
with $z$ from 0.1 to 1.2. Comparing the number of
galaxies with tidal structures with the total number of
objects in this $z$ range, they constructed the redshift
dependence of the fraction of interacting galaxies.
The derived dependence is described by the exponent
$m=2.25\pm0.24$; however, if the least reliable intervals
near the beginning and the end of the investigated
z range are eliminated, then $m=2.56\pm0.24$
(this value is given in Table 2). Note also that Bridge
et al. (2010) considered galaxies with stellar masses
$>10^{9.5}$\,M$_{\odot}$ (Table 2). Our approach to estimating
$m$ and the observational data used were completely
different, but the results turned out to be close.

The paper by Lavery et al. (2004) on collisional
ring galaxies gives a very large value of $m$. However,
the sample of galaxies they used is small. In addition,
the type of interaction itself in which one of the galaxies
passes through the center of the other at a high
velocity is so rare that the local space density of such
objects and, consequently, the value of $m$ are known
poorly.

The statistics of merging galaxies (mergers) also,
on the whole, is indicative of relatively large
($\approx 2-4$) $m$ (Table~2). The only exception is the paper by
Lotz et al. (2008); however, there are several factors
(e.g., a strong dependence of the result on the data
in two bins with the lowest $z$) that may explain this
discrepancy (for a discussion, see Bridge et al. 2010).
As a result, we may conclude that the works of
recent years give a relatively consistent picture of the
growth of the frequency of morphological signatures
of interactions and mergers up to $z \sim 1$.
Most of the results agree, within the error limits, with 
$m \approx 2.6$. The statistics of binary systems, including M51-type
galaxies (this paper), also yields similar results, $m \sim 3$
(Le F\`evre et al. 2000; Kartaltepe et al. 2007; de
Ravel et al. 2011).

The conversion from the observed fraction of interacting
and merging galaxies to the galaxy merger
rate ($R_{mg}$) is usually accomplished through the timescale 
$t_v$ on which the galaxies appear morphologically
peculiar (e.g., exhibit distinguishable tidal structures):
$R_{mg}(z) = \delta(z)/t_v$, where $\delta(z)$ 
is the fraction of interacting galaxies at redshift $z$. For galaxies
with tidal tails, we assumed that $\delta_0=0.02$;
at $m = 2.6$ for galaxies with $M(B)\leq -18^m$
(Table 2), we then obtain $\delta(z=0.7) = 0.08$.
As the time during which the galaxies exhibit noticeable tidal tails, we
take, according to Bridge et al. (2010), $t_v$=0.8 Gyr.
(Of course, this estimate is very uncertain, because
it depends, in particular, on the mass ratio of the
interacting galaxies, the gas fraction in them, the
method of identifying tidal structures, etc.) For the
galaxy merger rate at $z = 0.7$, we then obtain
$R_{mg} \approx 0.1$ mergers per galaxy during 1 Gyr. This estimate
agrees well with the data of other authors obtained by
different methods (see Fig. 11 in Bridge et al. 2010).
If we pass to a unit volume, then the merger rate is
$\sim 10^{-3}$/(Gyr $\times$ Mpc$^3$).

Using the above numbers, we can estimate the
averaged history of interactions for the galaxies of our
sample. Integrating the fraction of interacting galaxies
at different $z$ by taking into account the timescale
$t_v$ (see Eq. (3) in Bridge et al. 2010), we found that
a typical galaxy with $M(B)\leq -18^m$ 
in the $z$ range from 0.7 to 0.0 (a time of 6--7 Gyr 
corresponds to this range) underwent $\sim 0.35$
mergers or close encounters
accompanied by the formation of extended tidal tails.
In other words, a third of bright galaxies have undergone
strong gravitational perturbations and mergers
in the last 6--7 Gyr.

In the above simple estimates, there is one, previously
almost undiscussed uncertainty: the timescale
$t_v$ can depend on $z$. As was shown by Mohamed et al.
(2011), the tidal structures in distant galaxies, on
average, appear shorter than those in nearby objects.
One of the possible causes is observational selection
due to the cosmological brightness dimming and the
influence of the $k$-correction. As a result of this selection,
in distant galaxies we predominantly observe
a relatively early evolutionary stage of the tails, when
they have a high surface brightness, while in nearby
galaxies we see, on average, ``older'' and longer structures.

To quantitatively estimate the influence of this
effect on $R_{mg}$, we used the results of the calculations
by Mihos (1995). Mihos showed that the tidal
tails should be visible in the HST exposures only for
$\sim 150$ Myr at $z = 1$ and for about 350 Myr at $z = 0.4$.
Assuming that $t_v = 800$ Myr for $z = 0$ (see above),
we can obtain the following simple approximation for
the $z$ dependence of $t_v$:
 $t_v = a\,{\rm exp}(-z/b)+c$, where
$a = 0.7$, $b = 0.4$, $c = 0.095$, and the timescale $t_v$ is
in Gyr. For $z = 0.7$, it follows from this approximation
that $t_v = 0.22$ Gyr and, accordingly,
$R_{mg}\approx0.36$ mergers per galaxy during 1 Gyr. For the full
history of mergers between $z = 0.7$ and $z = 0.0$, we
obtain an estimate of $\sim$0.75, i.e., 3/4 of all galaxies
with $M(B) \leq -18^m$ must have undergone mergers
in the last 6-7 Gyr.

\subsection{M51-type galaxies}

Let us now consider M51-type galaxies. As we
showed in this paper, their space density at
$z \leq 0.7$ evolves with $m = 2.5 - 2.7$
(Table~1). Assuming that $m = 2.6$ and
$\delta_0=0.003$ (Klimanov 2003), we can
estimate that the relative fraction of such galaxies
is 0.012 at $z = 0.7$ and 0.014 at $z = 0.8$. Recently,
L\'opez-Sanjuan et al. (2011) presented the results of
direct counts of faint satellites (with luminosities from
0.1 to 0.25 of the luminosity of the central galaxy) for
various distances from the main galaxy for galaxies
with $M(B) \leq -20^m$ at $z = 0.5$ and $z = 0.8$ based on
the VIMOS VLT data (VVDS-Deep survey). Extrapolating
these counts to 15 kpc (most of the satellites
in our sample of M51-type galaxies are within
this distance), we can estimate that the fraction of such
galaxies at $z = 0.8$ is $\approx$0.02. Given the large errors in
such estimates (different methods of identifying objects,
different constraints on their luminosity, etc.),
the agreement between the estimates of the fraction of
galaxies with relatively low-mass satellites at $z = 0.8$
may be recognized to be satisfactory.

The satellites near the outer boundary of the stellar
disks in the main galaxies must be rapidly swallowed.
The characteristic merger timescale is $\sim0.2-0.4$
Gyr (see, e.g., Lotz et al., 2010). Consequently, the
merger rate estimate for M51-type galaxies at $z = 0.7$ is
$R_{mg} \approx 0.03-0.06$ per galaxy brighter than
$M(B)=-18^m$ in 1 Gyr. L\'opez-Sanjuan et al. (2011)
give a similar value: the minor merger rate at
$z=0.5-0.8$ is 0.0045-0.034 (Table~8 in their paper).
The total expected number of mergers per galaxy at
$z < 0.7$ is 0.1--0.2. Therefore, M51-type galaxies
make a noticeable and non-negligible (compared to
major mergers) contribution to the galaxy merger rate
at $z = 0.7$. Of course, this conclusion is very sensitive
to the satellite merger timescale. If, for instance,
it is close to 1 Gyr (Kitzbichler and White 2008), then
the merger rate through M51-type galaxies decreases
by several times.

\section{Conclusions}

Based on a large sample of distant galaxies with
tidal tails and M51-type galaxies (spiral galaxies with
a relatively low-mass satellite located near the end of
one of the spiral arms), we estimated the evolution
of the space densities of objects of these types up to
$z = 0.7$. It turned out that their observed densities
increases with redshift approximately as $(1 + z)^{2.6}$.

At $z = 0.7$, the merger rate leading to the formation
of extended tidal tails is $\approx0.1$ per galaxy brighter
than $M(B)=-18^m$ in 1 Gyr. The corresponding
merger rate for M51-type galaxies is approximately
a factor of 2--3 lower.

In the last 6--7 Gyr, i.e., at $z \leq 0.7$, about a third
of the galaxies with $M(B) \leq -18^m$ must have undergone
strong gravitational perturbations and mergers; 
$\sim$1/10--1/5 of the galaxies swallowed nearby
satellites with $Ls/Lm\approx0.1-0.2$. Such processes
are capable of radically changing the characteristics
of galaxies (Toomre, 1977) and stimulating the processes
of star formation and nonthermal activity of the
nuclei (see, e.g., Keel et al. 1985).

The estimates of the galaxy merger rate depend
strongly on the adopted timescale on which they
appear peculiar ($t_v$). For instance, allowance for the
possible redshift dependence of $t_v$ (the identification
time of tidal structures decreases with increasing $z$)
can increase the above merger rates by several times.

The high galaxy merger rates found from observations
of distant objects clearly suggest that gravitational
interactions and mergers were among the
most important processes that determined the individual
properties of the galaxies surrounding us.
Furthermore, they affect the evolution of the
``luminous'' matter in the Universe as a whole, because
mergers change the luminosity function of galaxies,
the luminosity density produced by them, and other
characteristics.

\bigskip
\section*{Acknowledgments}
This work was supported in part by the 
"Bourse de la Ville de Paris" programme.

\section*{REFERENCES}

\indent

1. C.M. Baugh and G. Efstathiou, Mon. Not. R. Astron.
Soc. 265, 145 (1993).

2. J. Bicker, U. Fritze-v. Alvensleben, et al., Astron.
Astrophys. 413, 37 (2004).

3. C.R. Bridge, P.N. Appleton, C.J. Conselice, et al.,
Astrophys. J. 659, 931 (2007).

4. C.R. Bridge, R.G. Carlberg, and M. Sullivan, Astrophys.
J. 709, 1067 (2010).

5. Ch.J. Conselice, M.A. Bershady, M. Dickinson, and
C. Papovich, Astron. J. 126, 1183 (2003).

6. Ch.J. Conselice, C. Yang, and A.F.L. Bluck, Mon.
Not. R. Astron. Soc. 394, 1956 (2009).

7. O. Fakhouri and Ch.-P. Ma, Mon. Not. R. Astron.
Soc. 386, 577 (2008).

8. A. Gabasch, R. Bender, S. Seitz, et al., Astron. Astrophys.
421, 41 (2004).

9. J.E. Hibbard and W.D. Vacca, Astron. J. 114, 1741
(1997).

10. P. Kampczyk, S.J. Lilly, C.M. Carollo, et al., Astrophys.
J. Suppl. Ser. 172, 329 (2007).

11. J.S. Kartaltepe, D.B. Sanders, N.Z. Scoville, et al.,
Astrophys. J. Suppl. Ser. 172, 320 (2007).

12. W.C. Keel, R.C. Kennicutt, E. Hummel, and
J.M. van der Hulst, Astron. J. 90, 708 (1985).

13. M.G. Kitzbichler and S.D.M. White, Mon. Not.
R. Astron. Soc. 391, 1489 (2008).

14. S.A. Klimanov, Astrofizika 46, 191 (2003).

15. S.A. Klimanov and V.P. Reshetnikov, Astron. Astrophys.
378, 428 (2001).

16. R.J. Lavery, A. Remijan, V. Charmandaris, et al.,
Astrophys. J. 612, 679 (2004).

17. O. Le F\`evre, R. Abraham, S. J. Lilly, et al.,Mon. Not.
R. Astron. Soc. 311, 565 (2000).

18. C. L\'opez-Sanjuan, M. Balcells, C.E. Garcia-Dabo,
et al., Astrophys. J. 694, 643 (2009a).

19. C. L\'opez-Sanjuan, M. Balcells, P.G. P\'erez-Ganz\'alez, 
et al., Astron. Astrophys. 501, 505 (2009b).

20. C. L\'opez-Sanjuan, O. Le F\`evre, L. de Ravel, et al.,
Astron. Astrophys. 530, A20 (2011).

21. J.M. Lotz, M. Davis, S.M. Faber, et al., Astrophys.
J. 672, 177 (2008).

22. J.M. Lotz, P. Jonsson, T.J. Cox, and J.R. Primack,
Mon. Not. R. Astron. Soc. 404, 575 (2010).

23. H.J. McCracken, M. Radovich, E. Bertin, et al.,
Astron. Astrophys. 410, 17 (2003).

24. N. Metcalfe, T. Shanks, A. Campos, et al., Mon. Not.
R. Astron. Soc. 323, 795 (2001).

25. J.Ch. Mihos, Astrophys. J. 438, L75 (1995).

26. S.H. Miller, K. Bundy, M. Sullivan, et al., Astrophys.
J. (2011, in press); arXiv:1102.3911v1.

27. A. Miskolczi, D.J. Bomans, and R.-J. Dettmar, 
Astron. Astrophys., (2011, in press);
arXiv:1102.2905v1.

28. H. Mo, F. van den Bosch, and S.D.M. White, Galaxy
Formation and Evolution (Cambridge Univ. Press,
Cambridge, 2010).

29. Y.H. Mohamed and V.P. Reshetnikov, Astrophysics
54, 155 (2011).

30. Y.H. Mohamed, V.P. Reshetnikov, and N.Ya. Sotnikova,
Astron. Lett. 37 (2011); arXiv:1108.6155v1.

31. P.B. Nair and R.G. Abraham, Astrophys. J. Suppl.
Ser. 186, 427 (2010).

32. P. Norberg, Sh. Cole, C.M. Baugh, et al., Mon. Not.
R. Astron. Soc. 336, 907 (2002).

33. L. de Ravel, P. Kampczyk, O. Le F\`evre, et al., Astron.
Astrophys. (2011, in press); arXiv:1104.5470v1.

34. V.P. Reshetnikov, Astron. Astrophys. 321, 749 (1997).

35. V.P. Reshetnikov, Astron. Lett. 26, 61 (2000a).

36. V.P. Reshetnikov, Astron. Astrophys. 353, 92 (2000b).

37. V.P. Reshetnikov and R.-Yu. Dettmar, Astron. Lett.
33, 222 (2007); arXiv:astro-ph/0703784v1.

38. A. Toomre, Evolution of Galaxies and Stellar Populations,
Ed. by B.M. Tinsley and R.B. Larson (Yale
Univ. Observ., New Haven, 1977).

39. S.D.M. White and M.J. Rees, Mon. Not. R. Astron.
Soc. 183, 341 (1978).

40. S.E. Zepf and D.C. Koo, Astrophys. J. 337, 34
(1989).

\end{document}